\documentclass{aa}

\usepackage[varg]{txfonts}
\usepackage{graphicx}
\usepackage{amsmath}
\usepackage{amssymb}
\usepackage{microtype}
\usepackage{booktabs}
\usepackage{multirow}
\usepackage{tablefootnote}
\usepackage{threeparttablex}
\usepackage{hyperref}
\usepackage{diagbox}
\usepackage{pdflscape}
\usepackage{bm}

\makeatletter
\renewcommand*\aa@pageof{, page \thepage{} of \pageref*{LastPage}}
\makeatother

\hypersetup{colorlinks=true, linkcolor=blue, citecolor=blue, urlcolor=blue}

\begin{document}

    \title{Series of core collapse numerical simulations (SOLANUM)}
    \subtitle{I: Modelling the early and late infall in a sun-like protostar}
    \titlerunning{SOLANUM I - Modelling the early and late infall in a sun-like protostar}
    \authorrunning{D. Navarro-Almaida et al.}
    \author{
    D. Navarro-Almaida\inst{1}\corrauth{dnavarro@cab.inta-csic.es} 
    \and 
    B. Commerçon\inst{2}\email{benoit.commercon@ens-lyon.fr}
    \and 
    U. Lebreuilly\inst{3}\email{ugo.lebreuilly@cea.fr} 
    \and 
    P. Hennebelle\inst{3}\email{patrick.hennebelle@cea.fr}
    \and 
    A. Ahmad\inst{2}\email{adnan-ali.ahmad@cnrs.fr}
    \and 
    A. Fuente\inst{1}\email{afuente@cab.inta-csic.es} 
    \and
    P. Riviere-Marichalar\inst{4}\email{p.riviere@oan.es}
    \and
    J. E. Pineda\inst{5}\email{jpineda@mpe.mpg.de}
    }
    \institute{
        Centro de Astrobiolog\'ia (CAB), CSIC-INTA, Ctra. de Torrej\'on a Ajalvir km 4, 28850, Torrej\'on de Ardoz, Madrid, Spain
        \and
        Universit\'e Lyon 1, ENS de Lyon, CNRS, CRAL, UMR 5574, Lyon, France
        \and
        Universit\'e Paris-Saclay, Universit\'e Paris Cit\'e, CEA, CNRS, AIM, 91191 Gif-sur-Yvette, France
        \and
        Observatorio Astron\'omico Nacional (OAN, IGN), Calle Alfonso XII, 3, 28014, Madrid, Spain
        \and
        Max-Planck-Institut f{\"u}r extraterrestrische Physik, Gie{\ss}enbachstra{\ss}e 1, 85748 Garching bei M{\"u}nchen, Germany
        }
    \date{}
    \abstract{Magnetohydrodynamical simulations and interferometric observations toward young stellar objects reveal the accretion of gas that may alter the physical properties and chemical composition of protostellar disks.}
    {Our goal is to constrain the physical history and properties of late infalling gas, and assess its potential to change the physical and chemical properties of the disk.}
    {We carried out a core collapse simulation of 5 M$_{\odot}$ of gas endowed with tracer and sink particles during $\sim 3.6\times 10^{5}$ yr. We analyzed the properties of the rotationally supported disk, and the origin and physical history of the gas from which it is initially formed. We also selected the tracer particles that describe the late infall to analyze their origin and physical history, and compare them to the information obtained from the disk.}
    {The final mass of the sink reaches 1.02 M$_\odot$, while the mass of the disk stays fairly constant around 0.02 M$_\odot$. Throughout its evolution, the disk radius increases up to $\sim 40$ au and its density is reduced by a factor of $\sim 8$. Filaments that channel gas from the envelope to the disk and sink appear as a result of the magnetic interchange instability. Tracer particles can identify late infalling gas, whose physical properties resemble that of observed streamers and are only accreted in the final snapshots of the simulation. The origin and physical history of the late infalling gas are different from those of the gas that initially forms the disk.}
    {Core collapse simulations predict the appearance of filaments that connect the envelope to the disk. The late accretion of gas through these channels has properties similar to those of streamers and brings gas to the disk with a different physical history that could translate into a different chemical composition.}
    \maketitle
    \nolinenumbers
    \section{Introduction}

        Low-mass star formation is currently pictured as the process in which prestellar cores, located in the filamentary networks that form molecular clouds, become locally gravitationally unstable and undergo gravitational collapse \citep[see, e.g.,][]{Andre2014, Pineda2023}. The initial stage of this collapse is isothermal, as gravitational energy can be freely radiated away. It is only when the gas becomes dense and optically thick enough that the gravitational energy is trapped, heating the gas and forming the so-called first Larson core or first hydrostatic core \citep[FHSC,][]{Larson1969}. If the FHSC reaches temperatures beyond $\sim 2000$ K, molecular hydrogen dissociates. This endothermic process triggers a second collapse that leads to the formation of the second Larson core. The Class 0 protostar is born. After that, the protostellar core continues to accrete the remnants of the first core and the surrounding envelope. The classification of the newly formed young stellar object (YSO) into more evolved objects, Class I-II-III, typically depends on properties of the observed spectral energy distribution (SED), reflecting the balance between the mass of the central object $M_{\star}$ and that of the envelope $M_{\rm env}$. For example, Class 0 objects are typically characterized observationally by a high ratio of submillimeter to bolometric luminosity, believed to indicate that they have accreted less than half of their final mass \citep[$M_{\rm env}\gg M_{\star}$,][]{Andre2000}. In the transition from Class 0 to Class I, Class II, and Class III, protostellar systems are usually classified by the slope of the SED $\alpha_{\rm IR}$ in the $2-25\ \mu{\rm m}$ range, again highly influenced by the balance between the radiation of the protostar, disk, and envelope. In order to constrain how the masses of the envelope and the disk evolve from Class 0 to Class I YSOs, several protostars at different evolutionary stages have been observed in submillimeter continuum at high resolution to disentangle the continuum emission of the disk from that of the envelope and study the evolution of their masses \citep{Jorgensen2009, Frimann2017, Andersen2019}. No significant trends with age were observed in these studies, and the time scales describing protostellar accretion and envelope dissipation are still uncertain \citep{Dunham2014}.

        With increasing computing power, numerical simulations have also addressed this issue. There are a multitude of studies investigating the collapse of protostellar cores and the subsequent formation of disks and outflows \citep[see, e.g.,][]{Machida2010, Commerçon2012a, Commerçon2012b, Tsukamoto2013, Tomida2015, Hincelin2016, Bate2018, Hennebelle2020, Kuffmeier2023, Commerçon2024, Lebreuilly2024b}. These studies provide insights into the physics at play in the observed object and include the computation of synthetic observable quantities to be cross-validated with observations of objects in different evolutionary stages. They comprise cloud and core-scale setups to simulate isolated dense cores, and molecular clouds resolved down to individual protostars that allow for the synthesis of stellar populations and the inclusion of environmental feedback \citep{Pelkonen2021, Kuffmeier2023, Lebreuilly2024b}. In fact, the environment in which a YSO develops is a decisive factor in characterizing its evolutionary stage. As it turns out, the accretion history of a protostellar object can substantially alter its apparent age according to its bolometric temperature and be classified as much younger than its dynamical age \citep{Kuffmeier2023}. According to cloud-scale simulations \citep[see, e.g.,][]{Kuffmeier2023, Huhn2025}, this is a characteristic of protostars that are ejected from their birth environment and move through the parent molecular cloud at later stages, and protostars that accrete envelope material beyond the initial collapse phase that was not initially gravitationally bound. These phenomena are collectively known as late accretion.

        Late accretion is becoming a crucial aspect in the context of dust grain growth and planetary formation. The high angular resolution observations with ALMA have shifted the paradigm of planet formation in protoplanetary disks, as they revealed a widespread presence of rings, gaps, and other substructures even in Class I \citep{Ohashi2023, Hsieh2025, Shoshi2025, Maureira2026} objects, implying a much earlier onset of planet formation than previously thought. Although the origin of these structures is still debated, numerical simulations have shown that magnetohydrodynamical and gravitational instabilities induced by infalling material from the envelope might promote their formation \citep[see, e.g.,][]{Hennebelle2017, Kuznetsova2022}. For instance, the density gradients that appear at the landing sites of the accreted gas may lead to the appearance of vortices and dust traps that would enhance grain coagulation and the formation of planetesimals. Late accretion has an observational realization in the so-called streamers. Streamers are velocity-coherent filamentary structures observed around YSOs at different evolutionary stages that are thought to connect and transport material from the surrounding cloud and envelope to the central object \citep{Pineda2023}. These structures are observed in dust continuum and molecular line emission. For instance, hints of accretion streamers were presented by \citet{LeGouellec2019} toward young Class 0 objects in the Serpens molecular cloud using dust continuum polarization observations. Molecular line emission also allowed the detection of these features in Class 0 YSOs: C$^{18}$O $2\rightarrow1$ \citep[][]{Chou2016, Frimann2017}, HC$_{3}$N $10\rightarrow9$, and C$_{2}$S $6_{7}-5_{6}$ \citep{Pineda2020, Taniguchi2024, ValdiviaMena2024} towards sources in Perseus. Streamers are also detected towards Class I sources: HL Tau \citep{Yen2019, Garufi2022}, IRS 63 \citep{SeguraCox2020, Podio2024}, and Per-emb-50 \citep{ValdiviaMena2022}. Finally, despite the faint envelope surrounding Class II YSOs, several streamers have been detected around them: DG Tau \citep{Garufi2022} and BHB2007 \citep{Alves2020}. The detection of these features at several stages in the star formation process suggests that many mechanisms might be involved in their appearance. MHD simulations \citep[see, e.g.,][]{Kuffmeier2017, Hennebelle2020} confirm the presence of streamers on core scales as channeled accretion from a gas reservoir through filaments. Alternatively, it has been shown that they may also appear as the result of interactions of the central object with its environment through Bondi-Hoyle accretion \citep{Padoan2025}, a binary companion, or a close encounter \citep{Vorobyov2020}. Similar observational features also appear in the context of the magnetic flux problem in star formation. As the protostar accretes the surrounding material, magnetic flux accumulates towards the central object. In this accumulation, the magnetic flux would be much higher than observed if no mechanisms hindering the effective drag of magnetic fields or removing magnetic flux were present. The magnetic interchange instability, a phenomenon that removes magnetic flux from the central object, has been shown to naturally produce filamentary, cavity, and ring-like structures in numerical simulations \citep[see, e.g.,][]{Zhao2011, Hennebelle2020, Machida2020, Machida2025, Mayer2025}. This mechanism was proposed to explain the presence of arc-like structures and streamer-like features observed towards YSOs by, for instance, \citet{Tokuda2023}, \citet{Fielder2024}, \citet{Tokuda2024}, and \citet{Tokuda2026}.
        
        Since streamers are thought to bring gas from the envelope to the central object, they may alter the chemistry of the disk because it potentially has a chemical composition different from the gas close to the YSO. This possibility was investigated in, for example, \citet{Taniguchi2024}, where they found a streamer connecting a gas reservoir with the Class 0 object Per-emb-2. The abundance ratio C$_{2}$S/HC$_{3}$N observed towards the streamer and the reservoir was then compared to a static chemical model that indicated the infall of chemically young gas to the protostellar source that would potentially alter its chemical composition. As mentioned above, the infalling gas from the streamer also introduces instabilities and shocks to the disk. According to the results presented by \citet{Podio2024}, the shocks at the streamer landing sites lead to a local release of chemical species into the gas phase that would otherwise be locked in ices. The offset gas-phase detection of SO$_{2}$, a shock tracer, towards Per-emb-50, a Class I object known to harbor a streamer, also signals local changes in the physical conditions induced by them \citep{ValdiviaMena2025}.

        The properties of streamers and their impact on the chemical composition of the disk have only recently been studied in detail. In this paper, we present the numerical setup of a 5 $M_{\odot}$ collapsing core simulation and analyze the evolution of the disk properties. We then compare them with the physical history of the late infalling gas to assess its potential to change the chemical composition of the disk. Finally, we test whether the observed properties of streamers are in agreement with the predicted mass, size, and accretion rate of the filamentary structures that appear in core collapse simulations.
    
    \section{Numerical simulation of core collapse}

        We used the adaptive mesh refinement (AMR) code \texttt{RAMSES} \citep{Teyssier2002} to compute the evolution of the physical properties of the collapsing core over time. The numerical setup is similar to the one used by \citet{Commerçon2024} and \citet{Ahmad2026}, and it is detailed below.
        
        \subsection{Simulated physics}
        
            The numerical framework used with \texttt{RAMSES} integrates the equations of radiation magnetohydrodynamics (R-MHD) in its non-ideal implementation. The non-ideal MHD solver only accounts for ambipolar diffusion \citep{Masson2012} as a non-ideal effect. Ambipolar diffusion resistivities were computed using the chemical network described in \citet{Marchand2016}, suited to the conditions characteristic of prestellar core collapse. Radiative transfer was performed following the hybrid irradiation approach developed by \citet{MignonRisse2020}. This hybrid scheme combines the M1 closure method for stellar radiation \citep{Levermore1984, Rosdahl2013, Rosdahl2015} and the grey flux-limited-diffusion (FLD) approximation \citep{Minerbo1978, Levermore1981, Commerçon2011, Commerçon2014} to handle radiation transport.

            \begin{figure*}
                \centering
                \includegraphics[width=0.87\textwidth]{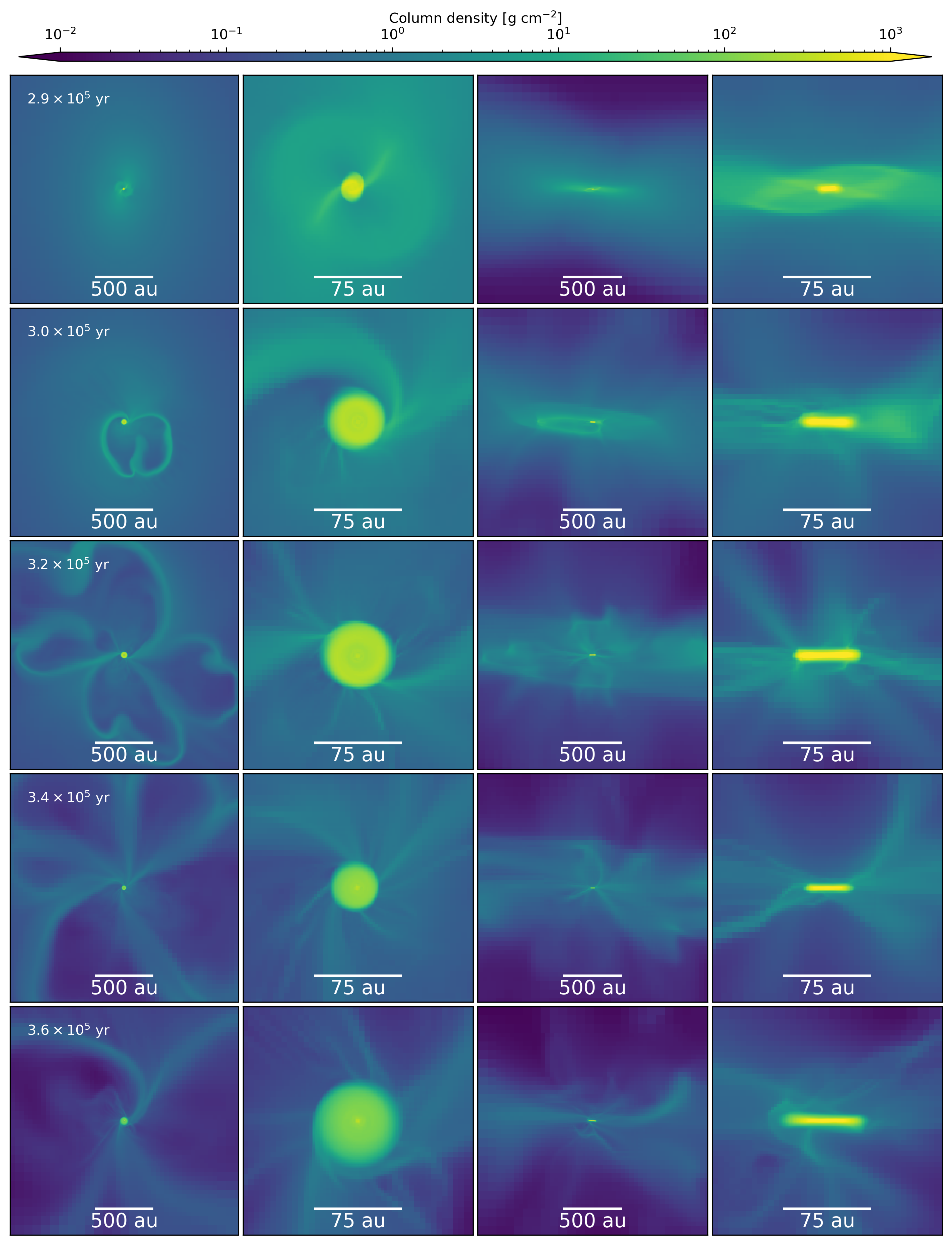}
                \caption{Column density maps at different times of the structures that develop in the core collapse simulation. Each row shows large-scale (first and third columns) and disk-scale (second and fourth columns) column density maps in face-on (first and second columns) and edge-on (third and fourth columns) views for a given snapshot.}
                \label{fig:snapshots}
            \end{figure*}
            
        \subsection{Initial conditions}

            \begin{figure*}
                \centering
                \includegraphics[width=\textwidth]{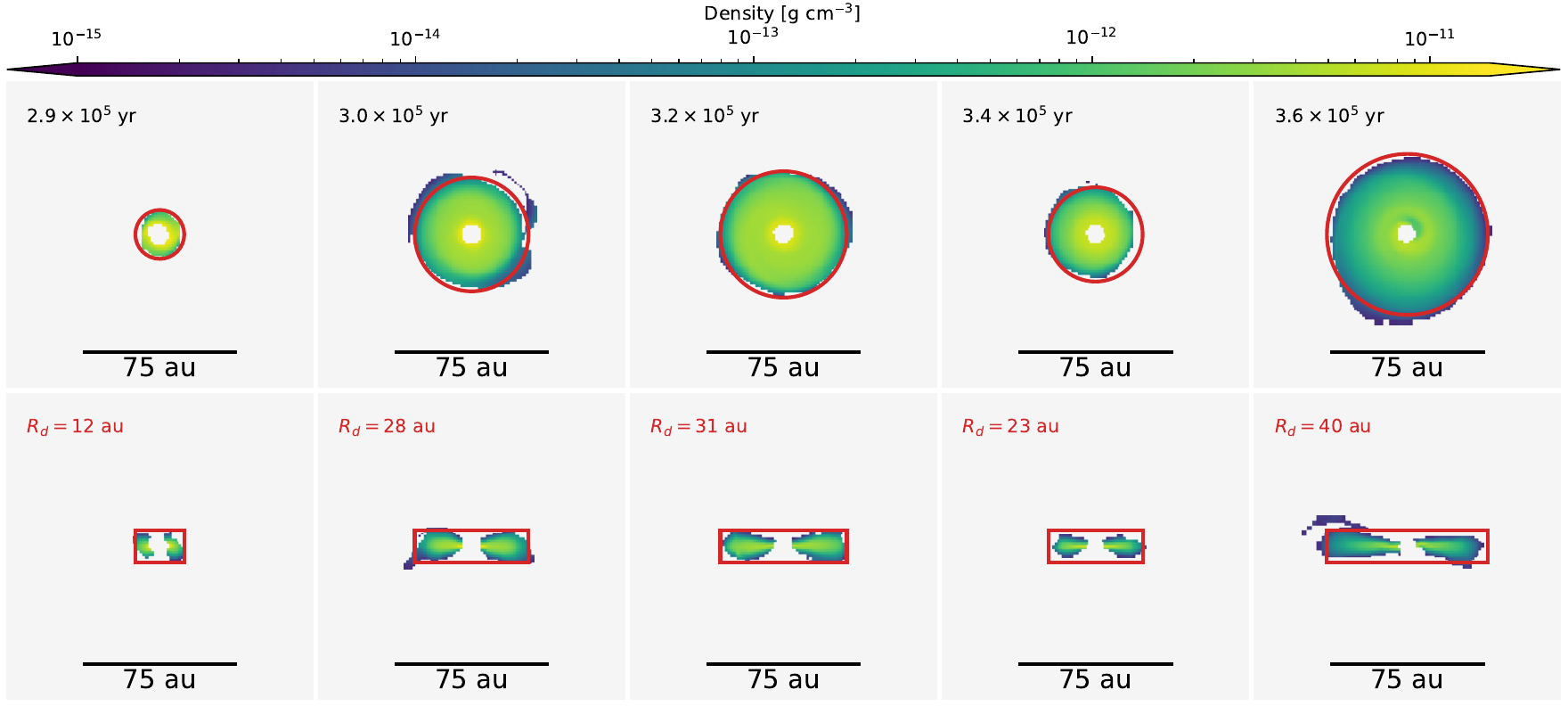}
                \caption{Properties of the rotationally supported disk in different snapshots. \emph{Top row:} each column shows the gas density slices of the rotationally supported disk in the face-on view of a given snapshot, from the moment of formation of the sink to the end of the simulation. The estimated size of the disk is shown as the radius of the red circle, centered in the sink particle. \emph{Bottom row:} edge-on views of the density slices of the disk in several snapshots. The red rectangle sets the vertical height of the disk and its width is the diameter of the red circle previously defined.}
                \label{fig:diskSnapshots}
            \end{figure*}
        
            Our simulation describes the collapse of an isolated dense core. The computational domain is a cube with a side length of $l\sim 7.9\times 10^{4}$ au and periodic boundaries. The size of the box is sufficiently large to ensure that the collapse evolution is not affected by the boundary conditions. The collapsing cloud has a total mass of 5 M$_{\odot}$ of gas evenly distributed in a sphere with an initial radius of $R\sim 2\times 10^{4}$ au and an initial uniform temperature of 10 K. These initial parameters resemble, for instance, the extent and temperatures of the moderate densities, $\lesssim 10^{5}$ cm$^{-3}$, observed towards the dense core Barnard 1b \citep[see, e.g.,][]{NavarroAlmaida2025}. The virial parameter $\alpha$, that is, the thermal-to-gravitational energy ratio, is set equal to $\alpha = 0.4$ in this simulation, and it is related to the mass $M$, size $R$, and temperature $T$ of the initial core such that
            \begin{equation*}
                \alpha=\frac{5}{2}\frac{R}{GM}\frac{k_{\rm B}T}{\mu_{\rm g} m_{\rm H}},
            \end{equation*}
            where $G$ and $k_{\rm B}$ are the gravitational and Boltzmann constants, respectively, $m_{\rm H}$ is the mass of the hydrogen atom, and $\mu_{\rm g} = 2.31$ is the mean molecular weight. The core is initially in solid body rotation, with a rotational-to-gravitational energy ratio $\beta$, set to $\beta = 0.04$. No turbulent velocity field was set. The chosen value is similar to that observed toward dense cores in dark clouds \citep[see, e.g.,][]{Goodman1993}. This parameter, often called ``rotational parameter'' \citep{Dib2010}, is related to the size $R$, mass $M$, and angular velocity $\omega$ in such a way that 
            \begin{equation*}
                \beta=\frac{1}{3}\frac{\omega^{2}R^{3}}{GM}.
            \end{equation*}
            The magnetic field strength corresponds to a ratio of the initial mass to flux ratio over the critical mass to flux ratio of 3. The magnetic field is initially misaligned with the rotation axis by ten degrees. 

            The minimum level of refinement in the mesh is $\ell_{\rm min} = 5$. The grid is then refined following the Jeans length criterion using 20 points per Jeans length up to the maximum level of refinement, set to $\ell_{\rm max} = 16$. This leads to a maximum spatial resolution of $\Delta x_{\rm max}\sim 1.22$ au. This simulation includes sink particles as a proxy for stars and their accretion of surrounding gas. A sink particle is created when the density threshold $n_{\rm thr} = 1.5\times 10^{13}$ cm$^{-3}$ is reached. When created, the sink particle accretes mass according to the density threshold criterion: only $10\%$ of the material inside the accretion radius $r_{\rm acc} = 4\times \Delta x_{\rm max}$ that is dense enough $n_{\rm acc} > n_{\rm thr}/3$ and is not thermally supported is accreted to the sink particle, increasing its mass. Our choice of the sink accretion threshold follows from previous works that constrain this value in order to allow rotationally supported disk formation for uniform collapsing cores \citep{Machida2014} and is supported by analytical models of young disk evolution \citep{Hennebelle2020}. Finally, to properly follow the flow of gas, we included Lagrangian tracer particles in the simulation. Initially, the $10^{6}$ tracer particles are distributed uniformly throughout the collapsing sphere of gas before collapsing. These massless particles follow the flow of the collapsing gas, registering the evolution of quantities such as density and temperature in each time step.
    
    \section{Results}

        The core collapse continued until the sink reached an age of $7\times10^{4}$ yr and a mass of 1.02 $M_{\odot}$.

        \subsection{General overview of the simulation \label{sect:overview}}

            To provide a general overview of the simulation, in Fig. \ref{fig:snapshots} we show the z-axis (first two columns) and x-axis (last two columns) views of column density at several time steps. As we show in the next subsection, the disk is contained in the xy plane and thus these panels offer face-on and edge-on views, respectively. Each row corresponds to different time steps in the simulation. Disk and large scale views are provided to show all the different components of the collapsing core: envelope, disk, and core. The first snapshot, $t = 2.9\times 10^{5}$ yr after the beginning of the collapse, is shown in the first row of Fig. \ref{fig:snapshots}. It is when the sink particle is created and the rotationally supported disk is born. The snapshot displays a small disk with radius $R_{d}\sim 21$ au and a rather elliptical shape produced by the two spiral arms that feed gas to it (see the definition of the disk in the next section). In the edge-on view, the outflow cavity walls have not yet been fully developed. The disk becomes larger in the next two snapshots, after $10^{4}$ yr and $3\times 10^{4}$ yr, respectively, with a radius $R_\mathrm{d}\sim 31$ au. Dense filaments connect the disk and the envelope, making the disk slightly eccentric and breaking the symmetry present in the first snapshot \citep{Ahmad2026}. This is a consequence of the magnetic interchange instability \citep{Krasnopolsky2012, Joos2012, Hennebelle2020, Machida2020, Machida2025}, which helps transport the magnetic flux outwards efficiently. The outflow cavity in these snapshots is more apparent by the characteristic conical shape \citep[see, e.g.][]{Bally2016}. In the next snapshot, after $5\times 10^{4}$ yr, the disk becomes smaller, with a radius $R_{d}\sim 26$ au. Finally, its radius increases again in the last snapshot, after $7\times 10^{4}$ yr, reaching $R_\mathrm{d}\sim 39$ au in a highly eccentric shape. Such eccentricities are not rare, as they are a natural outcome from the mode of accretion in MHD collapse \citep{Commerçon2024}. In the large-scale view of the sequence of snapshots, the envelope becomes progressively less denser as it is accreted into the disk and protostar. This is apparent in the decreasing column density of the envelope surrounding the central object.

        \subsection{Disk characterization and evolution of its properties}

            To select the disk structure in each snapshot, we applied the following criteria \citep{Joos2012, Hincelin2016, Hennebelle2020} to the cells in the AMR grid:
            \begin{itemize}
                \item Keplerian rotation is expected to occur in disks. To avoid rapid collapse in the radial direction, the azimuthal velocity $v_{\phi}$ is set to be two times higher than the radial velocity $v_{r}$, thus ensuring rotational support: $v_{\phi}>2\,v_{r}$.
                \item To prevent the inclusion of the outflow cavity and ensure hydrostatic equilibrium, the azimuthal velocity $v_{\phi}$ is set to be two times higher than any vertical motion: $v_{\phi}>2\,v_{z}$.
                \item To select the rotationally supported disk, we imposed the condition that the rotational support $e_\mathrm{rot} = \rho v_{\phi}^{2}/2$ dominates over thermal support, that is, the density of the internal energy $e_{i}$:
                \begin{equation*}
                    e_\mathrm{i} = \frac{1}{\gamma-1}\frac{\rho k_{B}T}{\mu_\mathrm{g}m_{\rm H}},\ \ e_{r} > 2 e_{i}. 
                \end{equation*}
                \item Density above the density threshold $10^{9}$ cm$^{-3}$. According to \citet{Joos2012}, this threshold provides realistic radii and heights for rotationally supported disks.
            \end{itemize}
            The positions and velocities for the application of these criteria were computed with the sink particle as the center. These criteria select a set of cells with a toroidal shape with an inner radius. To obtain the radius of the disk, we tracked the mass enclosed by a growing cylinder centered on the sink particle with an initial radius equal to the inner radius of the torus. As its radius progressively increases, so does the mass enclosed by the cylinder. The radius of the disk is then defined as the first radius from which the enclosed mass does not change more than $1\%$. This procedure is repeated for several disk heights also tracking the mass growth inside the cylinder, selecting the height that satisfies the same criteria as the radius. The set of cells corresponding to the disk and its estimated radius in a few snapshots are shown in Fig. \ref{fig:diskSnapshots}.
            
\begin{figure}
                \centering
                \includegraphics[width=0.49\textwidth]{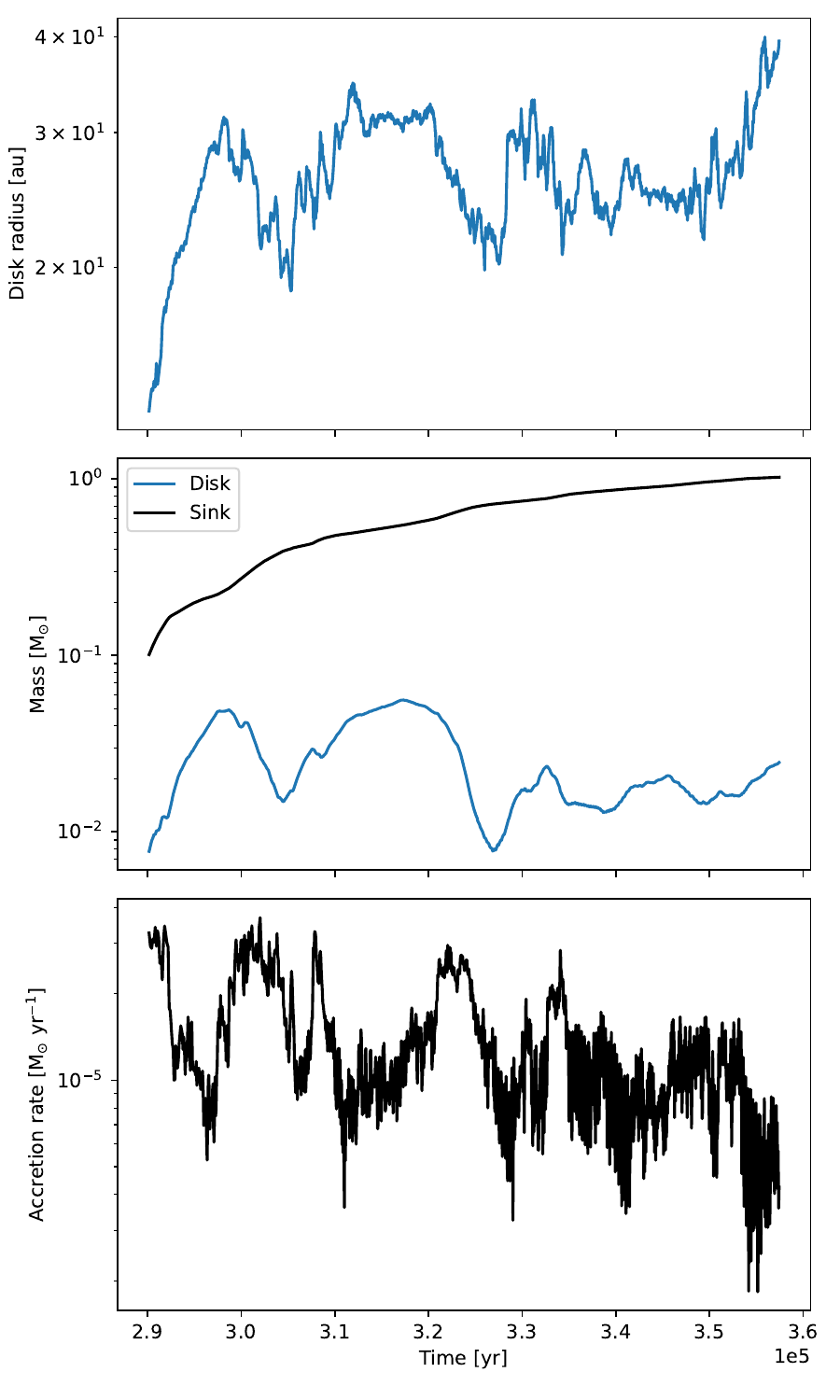}
                \caption{Disk and sink properties through time. \emph{Top panel:} the solid blue line describes the evolution of the disk radius. \emph{Middle panel:} the solid blue line shows the change in mass of the rotationally supported disk, while the solid black line corresponds to that of the sink particle. \emph{Bottom panel:} the solid black line describes the accretion rate to the sink in each time step.}
                \label{fig:diskSinkPropEvol}
            \end{figure}

            The disk is defined in each timestep as the set of cells in the AMR grid that follows the criteria described before. The properties of the disk, such as its radius, mass, and density structure, can be integrated over the set of cells that make up the disk. In Fig. \ref{fig:diskSinkPropEvol}, we show the evolution of the radius and mass of the disk in the top and middle panels, respectively. The sink mass and its accretion rate are also provided in the middle and bottom panels of Fig. \ref{fig:diskSinkPropEvol}, respectively. During the $\sim 7\times 10^{4}$ yr of evolution of the disk, there is a general increasing trend in the disk radius over time, reaching a size of up to $\sim 40$ au. The growth in the radius of the disk satisfies the analytical prediction in \citet{Hennebelle2016}, as its growth is proportional to $M_{\rm sink}^{1/3}$. With an age of $\sim 7\times 10^{4}$ yr, the disk in this simulation represents an early-stage protostellar disk. Its size is consistent with that of known disks around YSOs in several young star forming regions. For instance, most of the Class 0 sources in the CALYPSO sample \citep{Maury2019}, were best reproduced with disk-like continuum structures with radii $< 60$ au. Similar results were obtained in the CAMPOS survey, whose sample contains Class 0 and Class I objects \citep{Hsieh2024}. Most of the sources in this survey exhibit dust disk radii $< 50$ au. Radii in the range $11-34$ au were observed in the VANDAM survey \citep{SeguraCox2016}, which targeted Class 0 and Class I sources in the Perseus molecular cloud. The characteristic radius of young circumstellar disks observed in Lupus and Taurus, with radii $\sim 20$ au, also have comparable sizes \citep{Trapman2023}.

\begin{figure}
                \centering
                \includegraphics[width=0.49\textwidth]{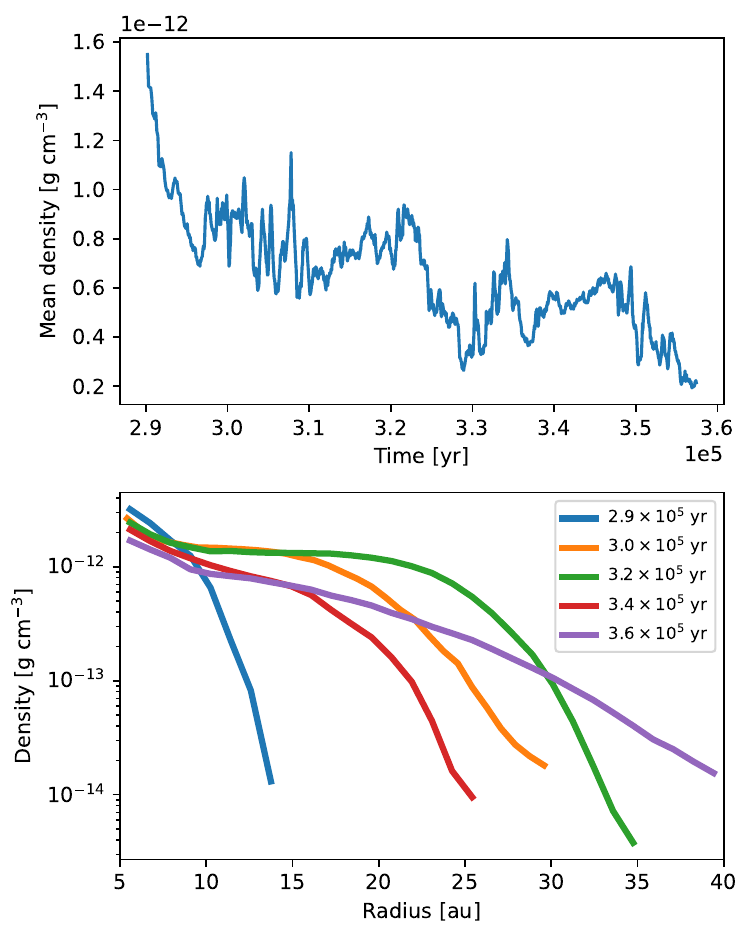}
                \caption{Gas density of the rotationally supported disk. \emph{Top panel:} evolution of the mean density of the disk, estimated as the ratio between its total mass and its total volume in each time step. \emph{Bottom panel:} the solid lines represent the azimuthally averaged density at several time steps.}
                \label{fig:meanDensity}
            \end{figure}

            The sink mass increases steadily until $1.02\ M_{\odot}$. However, the mass of the disk does not show any evolutionary trend and remains fairly constant around $M_\mathrm{d}\sim 0.02\ M_{\odot}$ despite its change in size. The thermally supported gas region surrounding the sink has, at most, a $6\%$ of the mass of the disk. In the set of simulations of \citet{Hennebelle2020}, a strong dependence of the disk mass on the density threshold of accretion to the sink $n_{\rm acc}$ was found. Since the value we chose for $n_{\rm acc}$ follows their prescription, our disk mass is comparable to that of their fiducial case $M_\mathrm{d}\sim 10^{-2}\ M_{\odot}$. The evolution of disk mass through Class 0 and Class I has been investigated observationally by, for instance, \citet{Jorgensen2009}, \citet{Frimann2017}, and \citet{Andersen2019}. By observing disks in different star-forming regions at high resolution, they found that there is no apparent correlation between the disk mass and the evolutionary stage, although estimates of the disk mass from the dust continuum are highly uncertain \citep{Dunham2014, Tung2024}. The accretion rate onto the sink is shown in the bottom panel of Fig. \ref{fig:diskSinkPropEvol}. Although the radius of the disk increases over time and the accretion to the sink decreases, this behavior does not translate into higher disk masses as seen in the middle panel of Fig. \ref{fig:diskSinkPropEvol}. This results in a decline of the gas density throughout the evolution of the disk. The change in mean density, that is, the total mass of the disk divided by its volume, is shown in the top panel of Fig. \ref{fig:meanDensity}. The mean disk density decreases about one order of magnitude during 70 kyr from $\sim 1.6\times 10^{-12}$ g cm$^{-3}$ to $\sim 2\times 10^{-13}$ g cm$^{-3}$. In the bottom panel of Fig. \ref{fig:meanDensity}, the azimuthally and vertically-averaged density at several time steps is shown. The density profile of the disk becomes shallower as time progresses, with its radius increasing (Fig. \ref{fig:diskSinkPropEvol}) and its peak density decreasing over time. The peak density starts below the density accretion threshold that we chose $n_{\rm acc} = 5\times 10^{12}$ cm$^{-3}$ 
            and is reduced by a factor of two at the end of the simulation.

            \subsection{Disk tracers}

            \begin{figure}
                \centering
                \includegraphics[width=0.49\textwidth]{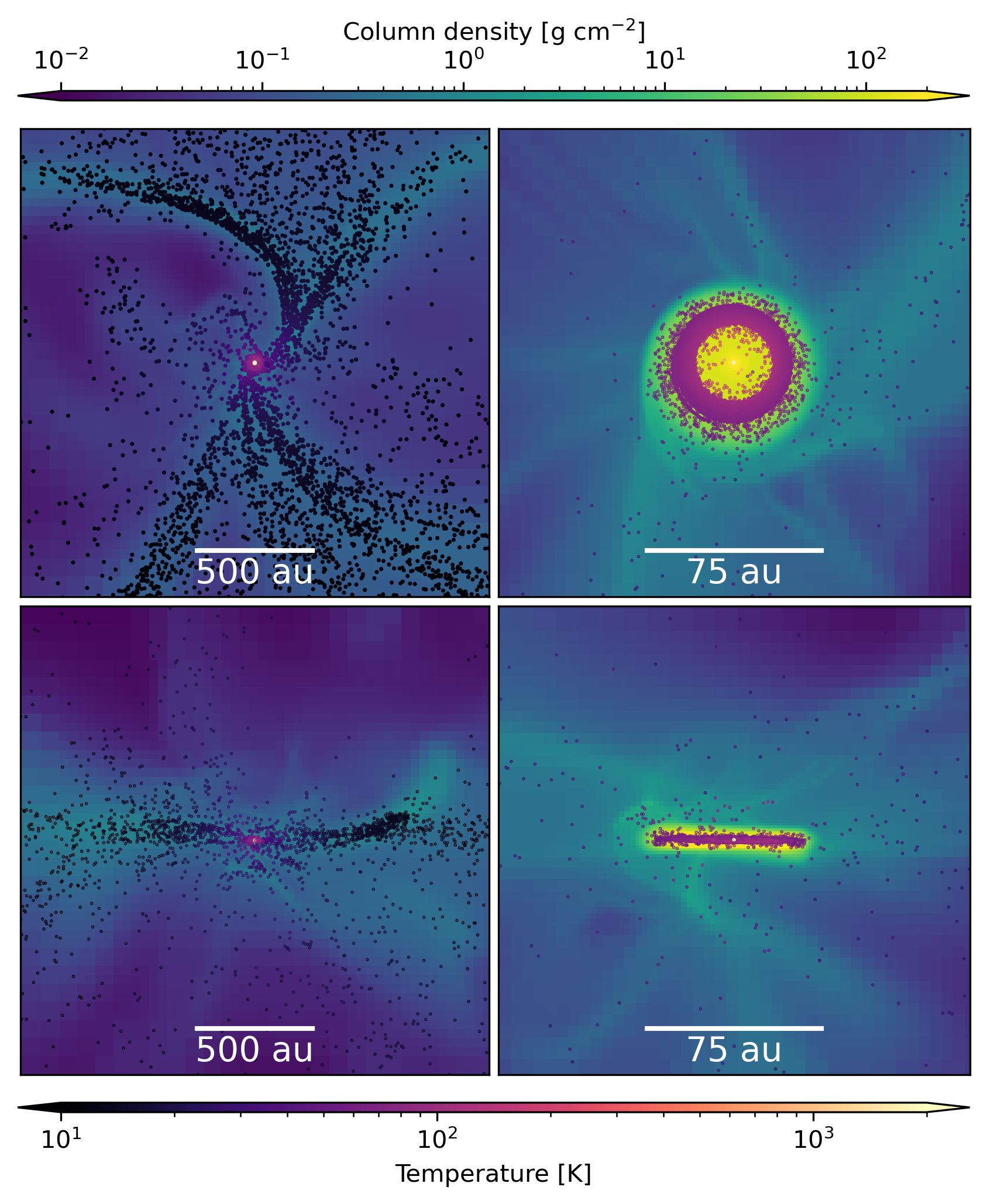}
                \caption{Tracer particles, disk tracers, and their properties at the end of the simulation. Only the projected positions onto the z plane of the tracer particles that satisfy $\left|z\right|<200$ au are shown. These panels display the distribution and temperature of the tracer particles (color bar at the bottom) that sample the protostellar environment. This environment is shown in face-on (top row) and edge-on (bottom row) column density views (color bar at the top). 
                }
                \label{fig:tracerSampling}
            \end{figure}
            
            The criteria over the density and rotational support to select the disk were also applied to the Lagrangian tracer particles at the end of the simulation. Figure  \ref{fig:tracerSampling} shows the density and temperature of the tracer particles that sample the disk and the surrounding areas. Among them, we selected the ones that satisfy the defining criteria of a rotationally supported disk above. These tracer particles can therefore be deemed disk tracers. Since tracer particles follow the flow of gas, they provide insight into the origin of the gas that eventually ends up forming the disk. Each tracer particle is endowed with a unique identification number, and thus can be tracked back in time. We tracked back in time the particles that form the disk in the final snapshot of the simulation, $7\times 10^{4}$ yr after the formation of the disk, and examined their physical properties at the time when the disk and sink are born.
            
\begin{figure}
                \centering
                \includegraphics[width=0.49\textwidth]{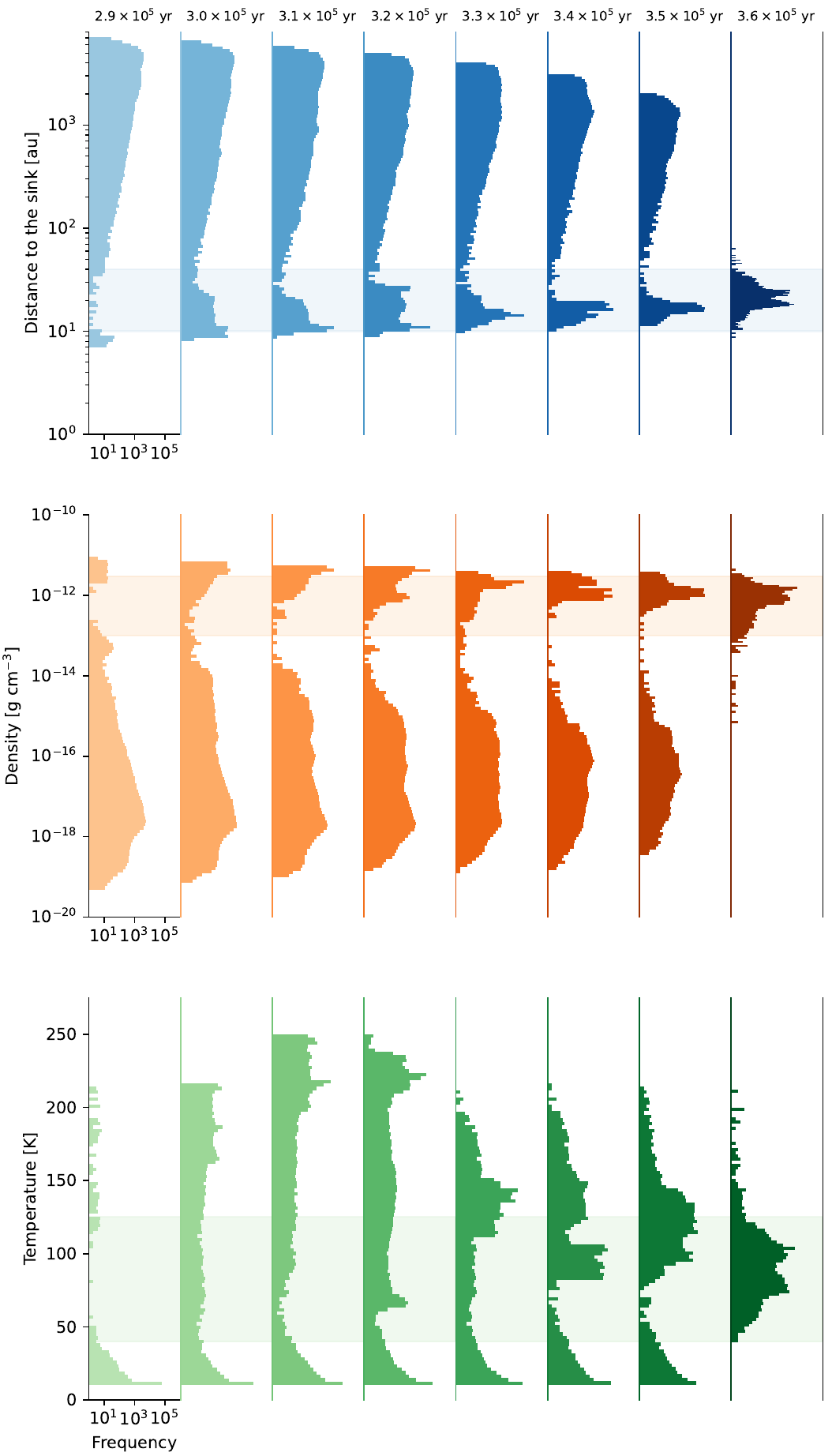}
                \caption{Distributions of physical properties of the tracer particles that form the disk at the end of the simulation in several time steps: distances to the sink (top panel), gas density (middle panel), and gas temperature (bottom panel).}
                \label{fig:diskTimeSeries}
            \end{figure}

            In Fig. \ref{fig:diskTimeSeries} we show how the distance, density, and temperature distributions of the tracer particles that form the disk at the end of the simulation evolve. In the top panel of Fig. \ref{fig:diskTimeSeries}, the distance from the sink is shown, revealing an upper bound to the distance: only the tracer particles, and therefore the gas, whose distance from the sink is lower than $\sim 7000$ au, end up forming the disk at the end of the simulation. Similarly, the lower bound is found at $\sim 3$ au. Initially, tracer particle distances are distributed in two groups: those who are already close to the sink ($\sim 10-20$ au away) and those who lie in the envelope, most of them at a distance of $3000-5000$ au. As accretion proceeds, tracer particles become confined at a distance to the sink in the range $10-30$ au, the size of the disk at the end of the simulation. The middle panel of Fig. \ref{fig:diskTimeSeries} shows the density distribution for the same set of tracer particles. They are again distributed in two groups: a low density one, with a wide dispersion $\sim10^{-19}-10^{-14}$ g cm$^{-3}$, and a high density one, with a low dispersion that peaks at $\sim 3\times 10^{-12}$ g cm$^{-3}$. Most of the particles are in the first group. However, as time evolves, the density of the tracer particles in the second group increases progressively, reaching $10^{-13}-10^{-12}$ g cm$^{-3}$ at the end of the simulation. The mean value of the gas in the disk calculated previously (top panel of Fig. \ref{fig:meanDensity}) is within this range. Finally, the evolution of the temperature distribution of the tracer particles that form the disk is shown in the bottom panel of Fig. \ref{fig:diskTimeSeries}. Initially, most tracer particles are at 10 K, although there is a fraction of them that are in the range $150-200$ K, belonging to the densest and closest sets of particles described previously. Tracer particles become increasingly warmer, depleting the 10 K peak and reaching $75-100$ K, the mean temperature of the disk at the end of the simulation. Interestingly, during their evolution, a significant fraction of tracer particles at 10 K reach even higher temperatures, of up to $200-250$ K after $3\times10^{4}$ yr, to then become colder again in the $75-100$ K range.

    \section{Discussion}

        In the following, we examine the properties of the gas channeled by the filamentary structures in the latest snapshots of the simulation as a realization of late accretion. Then, we assess the potential of this gas to rewrite the chemical composition of the disk. Finally, we test whether the predicted properties of the density enhancements and the gas they channel are in agreement with the observed properties of streamers.
        
        \subsection{Physical history of late infalling gas}\label{sec:physicalHistoLateInfall}

            The sink mass increases steadily in an approximately linear fashion (see Fig. \ref{fig:diskSinkPropEvol}) until the sink mass reaches $1.02\ M_{\odot}$ in the final snapshot of the simulation. This kind of mass growth has been shown to be typical of rather massive core collapse simulations \citep[$M_\mathrm{s}\geq 1 M_{\odot}$,][]{Kuffmeier2023}. Unlike a very low mass core collapse, where most of their mass is accreted during the initial collapse phase, in a more massive core collapse, a substantial amount of gas is expected to be accreted at later stages, even when this accreting mass was not initially bound to the prestellar core \citep[see, e.g.,][]{Smith2011, Pelkonen2021, Kuffmeier2023}. This process is commonly referred to as \emph{late accretion}.

\begin{figure*}
                \centering
                \includegraphics[width=0.49\textwidth]{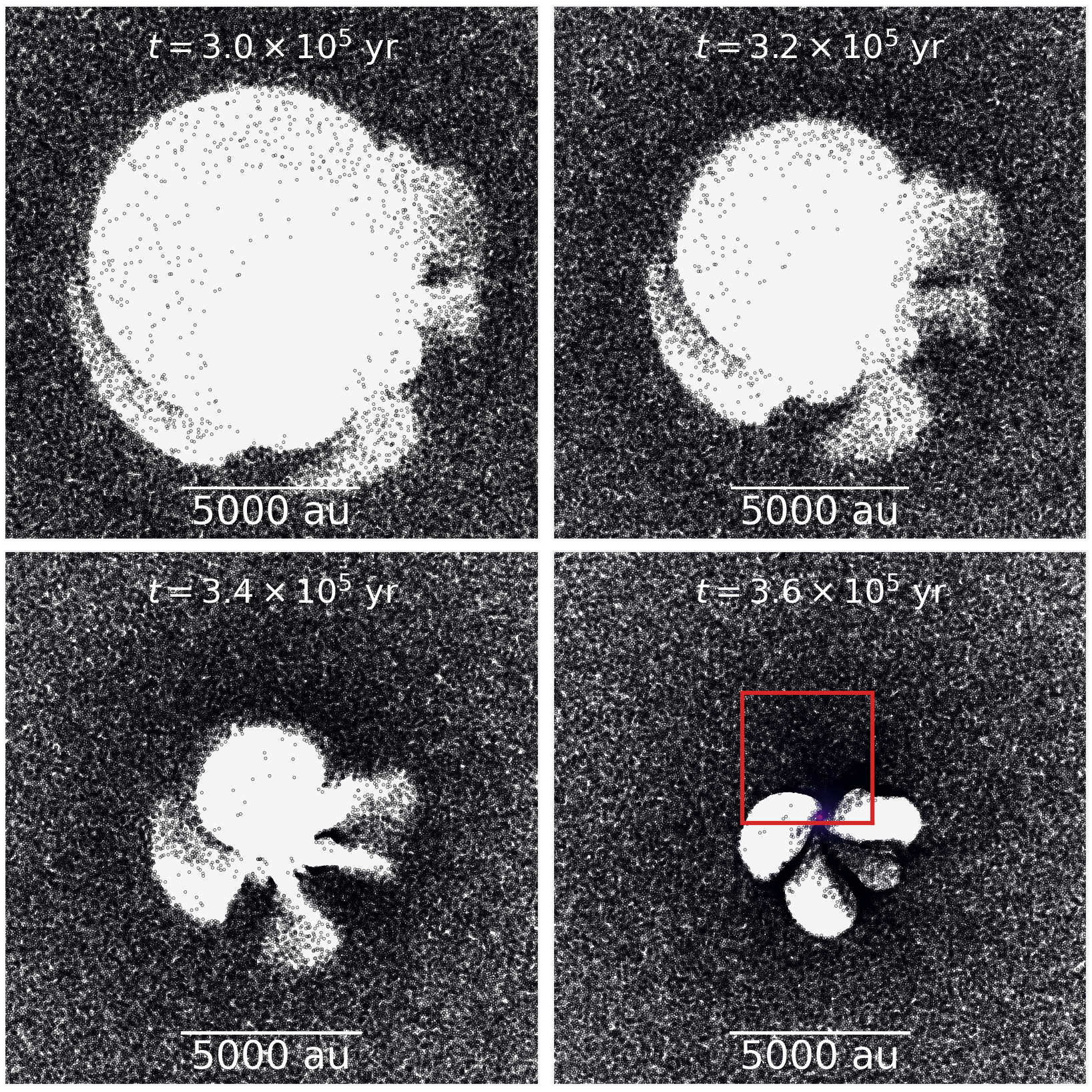}
                \includegraphics[width=0.49\textwidth]{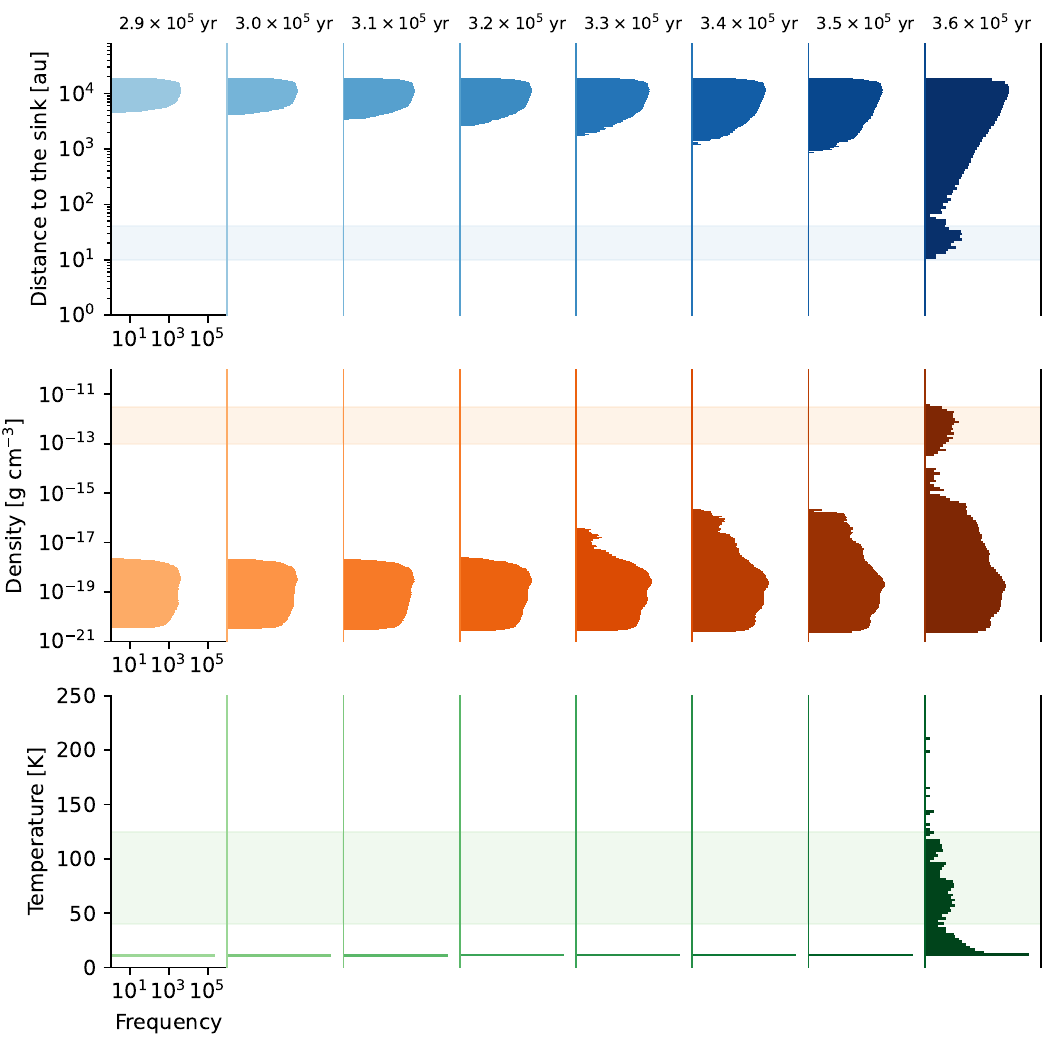}
                \caption{Infalling tracer particles and their physical history. \emph{Left:} face-on view of the infalling tracer particles at different time steps. A streamer is enclosed in the red square. \emph{Right:} distributions of physical properties of the infalling tracer particles: distances to the sink (top panel), gas density (middle panel), and gas temperature (bottom panel). The shaded ranges are the same as in Fig. \ref{fig:diskTimeSeries}.}
                \label{fig:infallTimeSteps}
            \end{figure*}

            Late accretion is of great interest in disk and planet formation in the context of the disk mass budget problem \citep{Greaves2010, Najita2014, Mulders2015, Mulders2018}, and as a potential replenisher of gas and dust around the disk. This gas potentially carries a chemical composition different from that of the disk. As mentioned above, this possibility was investigated by \citet{Taniguchi2024} towards the streamer found in Per-emb-2. Some of the proposed factors that could cause the chemical composition of streamers to be different from that of the disk are the origin and physical history of the streamer gas. We investigated this subject by examining the properties and physical history of the gas channeled from the outer envelope to the disk through the dense filaments present in our simulation. The tracer particles in our simulation allowed us to find and track this gas. To consider the gas that is infalling, we selected the tracer particles that satisfy the following condition at all time steps: $\vec{r}\cdot\vec{v} < 0$, where $\vec{r}$ is the position vector of the tracer particle and $\vec{v}$ its velocity. This condition selects $\sim 2.5\times 10^5$ tracer particles, one fourth of the total, whose spatial distribution at several time steps is shown on the left side of Fig. \ref{fig:infallTimeSteps}. This figure depicts the evolution of their spatial distribution at different time steps from the outer envelope to their accretion onto the disk. This condition not only selects the tracer particles that are always infalling, but also selects those that are accreted onto the disk in the latest snapshots of the simulation through the dense filaments formed during the collapse, becoming a realization of late accretion relative to the time span of the simulation. 
            
            This selection offers the opportunity to check whether the gas that is accreted onto the disk in the final snapshots of the simulation has a different physical and thermal history from the gas that is already contained in it (see Fig. \ref{fig:diskTimeSeries}) and thus can bring a potentially different chemical content to the disk. As we did in Fig. \ref{fig:diskTimeSeries}, we tracked the physical properties of the tracer particles to which we applied the condition above $\vec{r}\cdot\vec{v} < 0$. In the right panel of Fig. \ref{fig:infallTimeSteps}, we show the distributions of physical properties of the tracer particles to which we imposed the above condition ($\vec{r}\cdot\vec{v} < 0$). Infalling tracer particles come from a range of distances to the sink $(5-20) \times 10^{3}$ au that is generally beyond the distance of origin of the tracer particles that form the disk. If the chemical composition of the parent core is layered, the gas followed by these tracer particles would therefore have a different initial chemical composition to that that is already in the disk. Furthermore, the fraction of these tracer particles that reach the disk (shaded bands on the right side of Fig. \ref{fig:infallTimeSteps}) has, on average, a different thermal history: while a significant fraction of the tracer particles that form the disk undergo an early warming event $2\times 10^{4}$ yr after the formation of the disk, when the gas is heated to $\sim 250$ K to then cool and reach the mean temperature of the disk $\sim 100$ K, this does not occur for the late infalling tracer particles that end up in the disk, most of which are at 10 K until they are accreted onto the disk. The different origin and physical history of the gas accreted through the filaments, compared to the one that already forms the disk, become important factors suggesting that this accreting gas may alter the chemical composition of the disk.      

        \subsection{Can these filamentary structures account for the streamers observed towards YSOs?}\label{sec:comparisonObservations}
        
\begin{figure}
                \centering
                \includegraphics[width=0.49\textwidth]{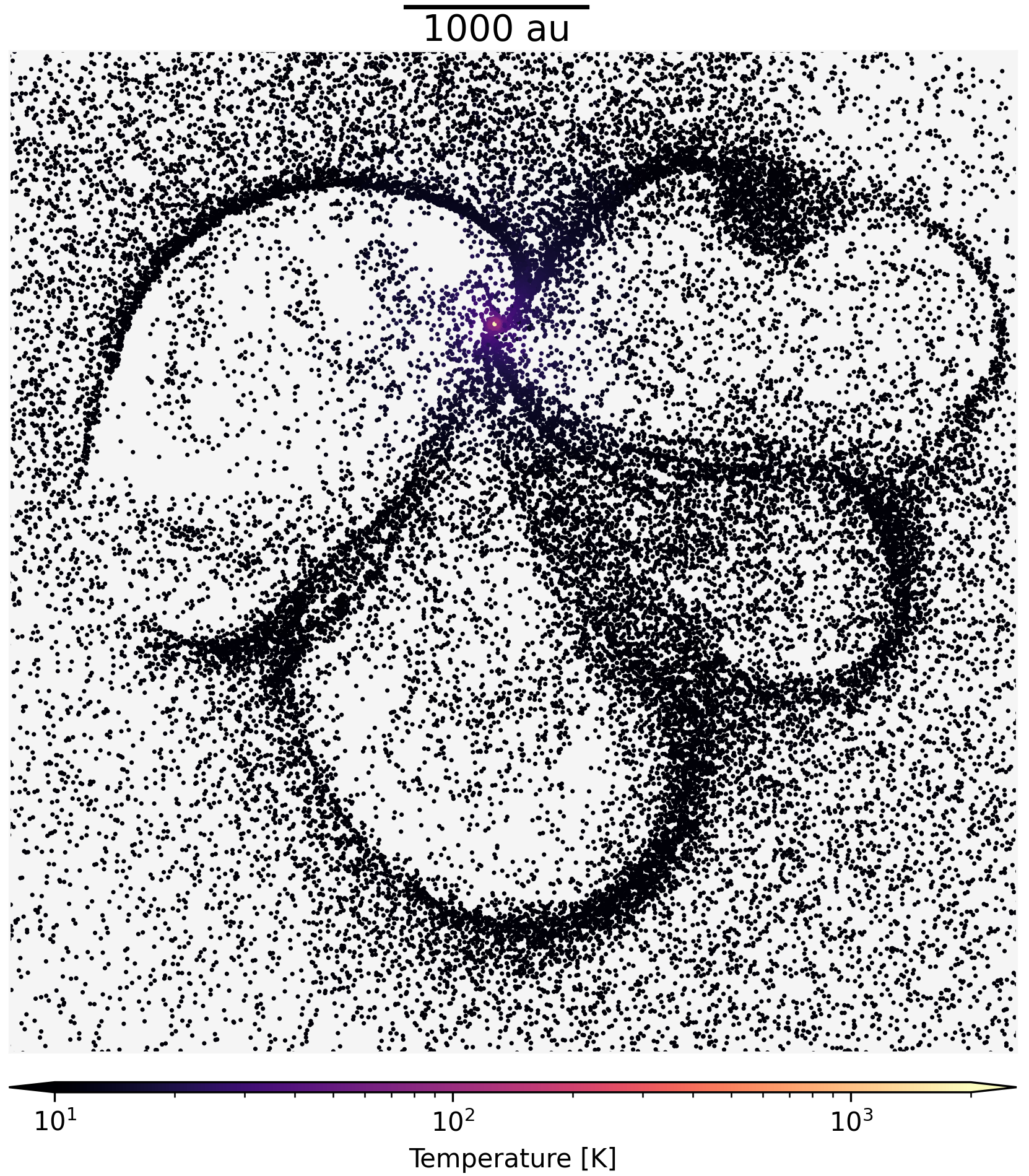}
                \caption{Spatial distribution of the tracer particles projected onto the z plane in the final snapshot of the simulation, displaying the bubble-shaped structures developed by the magnetic interchange instability.}
                \label{fig:bubbles}
            \end{figure}

            Late accretion has an observational realization in the so-called accretion streamers. As mentioned above, streamers are filamentary structures of multiple scales that have been observed with several chemical tracers in a growing body of young stellar objects \citep[see, e.g.,][]{Chou2016, LeGouellec2019, Yen2019, Alves2020, Pineda2020, Garufi2022, ValdiviaMena2022, Huang2023, ValdiviaMena2024}. Streamers are found in a wide variety of sizes, ranging from small-scale streamers of $\sim 10$ au wide to large-scale streamers of $\sim 1000$ au wide, and incident angles. This diversity reflects the still debated nature of late accretion and the different origin and mechanisms responsible for their appearance. In our simulation, filamentary enhancements are found in, for instance, the last snapshot (top-left panel of Fig. \ref{fig:tracerSampling}), channeling gas that is accreted onto the disk-sink system and is highly populated by tracer particles. To find the origin of this enhancement, Fig. \ref{fig:bubbles} shows a zoom out of the face-on view in the top-left panel of Fig. \ref{fig:tracerSampling}. In Fig. \ref{fig:bubbles}, it is apparent that the density enhancements that feed gas to the disk and sink are actually the walls of bubble-shaped structures developed during collapse. As discussed in Sect. \ref{sect:overview}, these bubbles are a consequence of the magnetic interchange instability that develops during accretion, compressing the surrounding gas into filamentary structures that feed gas from the envelope onto the disk and sink. The magnetic interchange instability and the morphology of these filaments may depend on numerical resolution \citep[see, e.g.,][]{Spruit1990}. However, the signatures of this phenomenon in terms of, for instance, anisotropic infall, are ubiquitous and found in many works that use different numerical methods and resolutions \citep[see, e.g., ][]{Machida2020, Hennebelle2020, Lee2021, Ahmad2026}.
            
            This mechanism is not expected to induce the formation of streamers in all stages of star formation. Although similar filamentary structures are detected in Class II sources, the magnetic interchange instability relies on the magnetic flux transport that occurs when the envelope is being accreted. In the case of Class II sources, a significant fraction of the envelope is already accreted, preventing the magnetic interchange instability from operating efficiently. Still, similar observational features as the ones predicted in this simulation are present in the literature. Magnetic-flux transport events like the magnetic interchange instability were proposed to produce the ring and arc-like structures reported by \citet{Tokuda2024, Tokuda2026} toward the disk around the embedded Class 0 protostar MC 27/L1521F in Taurus. However, several mechanisms such as fly-by encounters \citep{Kimmig2026}, cloudlet capture, and Bondi-Hoyle accretion \citep{Huhn2025} could also lead to the formation of streamers. Apart from the morphology, another property to compare with observations is the incident angle, that is, the angle between the streamer and the plane of the disk. This angle is estimated in the literature in a few cases using streamline models \citep{Pineda2020}. These streamline models compute the initial polar angle $\theta_{0}$, being the angle between the streamer and the z-axis, chosen to be in the direction of the outflow. We obtained the incident angle as $90-\theta_{0}$ for several streamers reported in the literature, which are collected in Table \ref{table:incidentAngles}. To calculate the incident angle in our simulation, we considered the set of infalling tracer particles shown in Fig. \ref{fig:infallTimeSteps}. In Fig. \ref{fig:incidentAngle}, we show the best-fit plane that contains the infalling tracer particles, whose normal vector is at an angle of $\sim 20^{\circ}$ with respect to the $z=0$ plane and contains the midplane of the disk. Although our prediction is in agreement with the streamer observed towards Per-emb 50, this does not necessarily imply that the streamer towards this source was formed by the magnetic interchange instability. It is expected that the chemical tracer used in the detection of the streamer also introduces uncertainties in the estimation of the angle of incidence.

\begin{figure}
                \centering
                \includegraphics[width=0.49\textwidth]{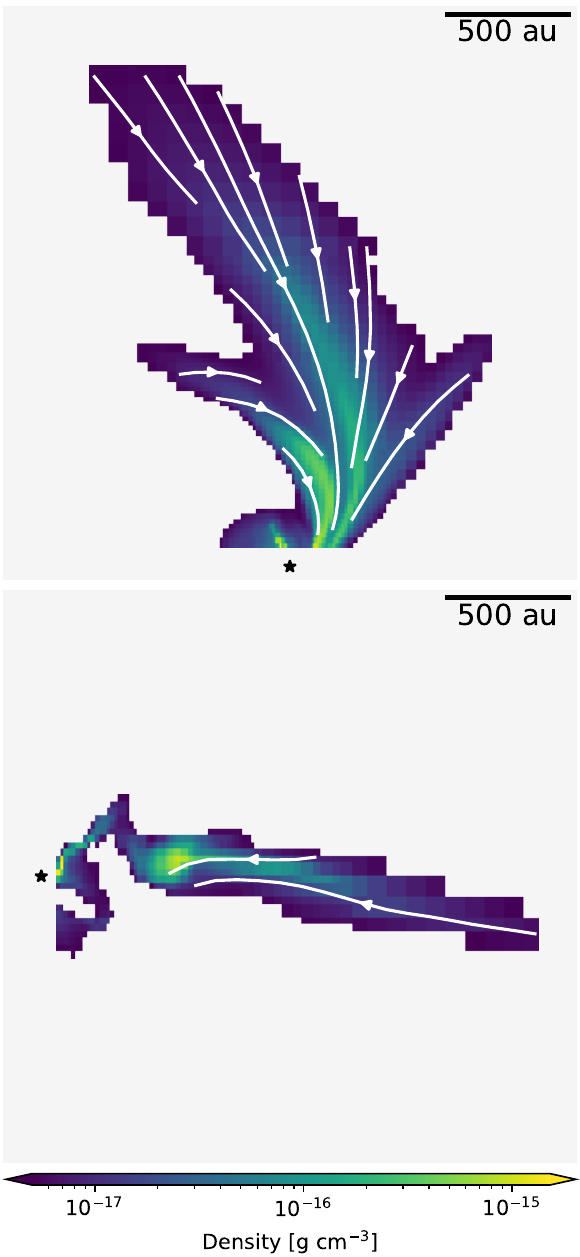}
                \caption{Face-on and edge-on slices of the cells that satisfy the critical density criterion $n > n_{\rm crit}$ inside the red square of Fig. \ref{fig:infallTimeSteps}. White arrows depict the gas flow.}
                \label{fig:nCrit}
            \end{figure}

            These structures channel gas from the envelope to the disk and, as we have seen, could carry a chemical composition different from that of the disk due to its origin and physical history. Therefore, they may be regarded as streamers. The density structure of streamers is poorly constrained in the literature. This difficulty likely arises from the lack of multi-transitional studies of streamers and the near local thermodynamic equilibrium (LTE) conditions of these density enhancements \citep[see, e.g.,][]{Tanious2025}. Furthermore, the observed size and morphology of streamers depend on the chemical tracer used. Observations of streamers are limited by the critical density of the observed rotational transition. For instance, the critical density of the DCN $3\rightarrow2$ line \citep{Gieser2025} at 10 K in the optically thin limit is $n_{\rm crit}\sim 3.4\times 10^{6}$ cm$^{-3}$ \citep{NavarroAlmaida2023}, a density threshold that would limit the extent of the observed infalling gas in our simulation, where most of it has a density of $10^{-19}$ g cm$^{-3}$ (right side of Fig. \ref{fig:infallTimeSteps}). In Fig. \ref{fig:nCrit} we show the $z=0$ and $y=0$ density slices of the gas density after applying the DCN $3\rightarrow2$ critical density cutoff $(n>n_{\rm crit})$ to the gas enclosed in the red square in the bottom-right panel of Fig. \ref{fig:infallTimeSteps}. The resulting filament is $R\sim 2500$ au wide. Similar sizes are found in streamers towards L1448 IRS3C \citep{Gieser2025}, using the same chemical tracer, and the Class I YSO Per-emb-50 \citep{ValdiviaMena2022}, observed with the p-H$_{2}$CO $3_{0,3}\rightarrow2_{0,2}$ rotational line, which has a similar critical density $n_{\rm crit}(10\ {\rm K})\sim 2.4\times 10^{6}$ cm$^{-3}$. Given the size of $r\sim 2500$ au, we estimated the mass of the streamer by adding the mass of cells that satisfy the critical density threshold (excluding the disk). The resulting mass of the streamer is $M_{\rm str} = 0.03\ M_{\odot}$, similar to that of other streamers in the literature: Per-emb-50 \citep{ValdiviaMena2022}, IRS3A \citep{Gieser2025}, and L1489 IRS \citep{Tanious2025}. With these estimations of size and mass, we finally computed the mean infall rate $\langle \dot{M}_{\rm in}\rangle$ as the mass of the streamer $M_{\rm str}$ divided by the free-fall time $t_{\rm ff}$:
            \begin{equation*}
                t_{\rm ff} = \sqrt{\frac{r^{3}}{GM_{\rm tot}}},
            \end{equation*}
            where $r$ is the size of the streamer, $G$ is the gravitational constant, and $M_{\rm tot}$ is the total mass enclosed within the distance $r$. Thus, the mean infall rate results in $\langle \dot{M}_{\rm in}\rangle = 1.72\times 10^{-6}\ M_{\odot}$ yr$^{-1}$. This rate is similar to the reported infall rates of the observed streamers in the literature above. The infall rate through this filament is approximately one third of the accretion rate to the sink. If the infalling material were to land on the disk, it would replace its gas in $t\sim 1.2\times 10^{4}$ yr, provided the infall rate remains constant. Since the infalling material has a very distinct physical history and origin, it has the potential to rewrite the chemical composition of the disk with which it was initially formed.  

    \section{Conclusions}

        We performed a 3D MHD simulation of a collapsing core of 5 M$_{\odot}$ of gas to investigate the properties and evolution of the gas that forms the disk and those of the infalling material at the end of the simulation. We modeled a streamer as part of the late infalling material and assessed its potential to bring a different chemical composition to the gas with which the disk was initially born. We concluded that:
        \begin{itemize}
            \item The disk is formed by gas coming from the inner $\sim 7000$ au of the collapsing core. As it accretes mass during its evolution, its density profile flattens, becoming larger while keeping its mass constant on average.
            \item Core collapse simulations predict the presence of filamentary structures that connect and transport gas from the surrounding envelope to the disk. In our simulation, these structures emerge because of the magnetic interchange instability.
            \item The properties of the predicted filamentary features resemble those of streamers detected toward YSOs. We estimated that their size is about several $\sim 1000$ au, a value dependent on the chemical tracer used in its detection. We also determined an average infall rate of $\langle \dot{M}_{\rm in}\rangle \sim 10^{-6}$ M$_{\odot}$ yr$^{-1}$ at the end of the simulation. These estimations of size and infall rate are in agreement with observations. 
            \item The gas outside the inner $\sim 7000$ au can be characterized as late infalling gas. It is accreted to the disk in the latest snapshots of the simulation through the filamentary structures described before. Its physical history exhibits significant differences in terms of temperature and density compared to the gas that initially forms the disk. The different origin and physical history of this gas suggest that it can bring a different chemical composition to the disk and alter its chemistry.
        \end{itemize}

        With the aid of numerical simulations, we investigated whether the filamentary structures that form in core collapse simulations can be identified with the streamers commonly seen toward YSOs. We characterized the properties of these structures and the physical history of the gas they channel to the disk, concluding that late infall can indeed rewrite the chemical history of the disk.

    \begin{acknowledgements}
        This work is supported by ERC grant SUL4LIFE, GA No. 101096293. Funded by the European Union. Views and opinions expressed are, however, those of the author(s) only and do not necessarily reflect those of the European Union or the European Research Council Executive Agency. Neither the European Union nor the granting authority can be held responsible for them. DNA also acknowledges funding support from the Fundaci\'on Ram\'on Areces through its international postdoc grant program. DNA and AF gratefully acknowledge the assistance of the IT team at CAB in setting up the computational resources used in this paper. BC and AAA acknowledge support from the French Agence Nationale de la Recherche (ANR) through the project PROMETHEE (ANR-22-CE31-0020). JEP was supported by the Max-Planck Society. AF and PRM thank project PID2022-137980NB-I00 funded by the Spanish Ministry of Science and Innovation/State Agency of Research MCIN/AEI/10.13039/501100011033 and by “ERDF A way of making Europe”.

    \end{acknowledgements}
    
    \bibliographystyle{aa}
    \bibliography{main.bib}

    \begin{appendix}
        \section{Incident angle and comparison with observations}
\begin{figure}[h]
                \centering
                \includegraphics[width=0.49\textwidth]{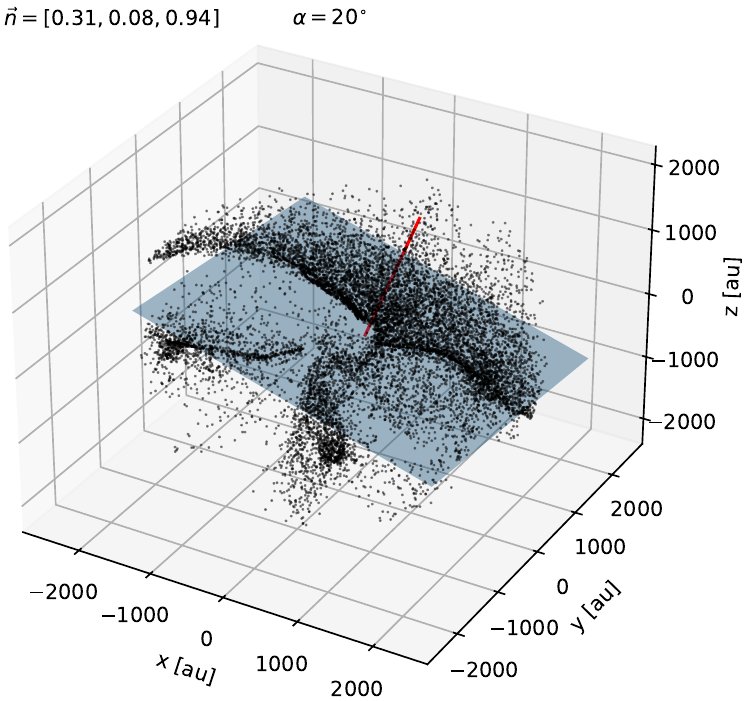}
                \caption{Infalling tracer particles belonging to late infall in the final snapshot of the simulation (black dots). The best-fit plane containing them is shown in blue and its normal vector is shown in  red.}
                \label{fig:incidentAngle}
        \end{figure}

        In this section we compute the incident angle of the filamentary structures through which late accretion gas proceeds in the final snapshot of the simulation as shown in Fig. \ref{fig:infallTimeSteps}. To do so, we computed the best-fit plane that contains the tracer particles satisfying the infalling criterion described in Sect. \ref{sec:physicalHistoLateInfall} (see Fig. \ref{fig:incidentAngle}). We excluded from this computation the tracer particles that are already accreted onto the disk and those that are further away from the streamer size derived in Sect. \ref{sec:comparisonObservations}. The incident angle was then obtained as the angle between the normal vector of the best-fit plane and that of the plane containing the disk. This results in an angle $\sim 20^{\circ}$. We also collected estimations of this angle in streamers observed in the literature using streamline models in Table \ref{table:incidentAngles}.
        
        \begin{table}[h]
            \caption{Collection of incident angles from the literature}
                \begin{threeparttable}
                    \resizebox{0.49\textwidth}{!}{
                    \begin{tabular}{lccc}
                        \toprule
                        \textbf{Source} & $\theta_{0}$ & $90-\theta_{0}$ & Reference \\ 
                        \midrule
                        This work   & $-$ & $20^{\circ}$ & $-$ \\
                        Per-emb 2   & $130^{\circ}$ & $-40^{\circ}$ &  \citet{Pineda2020}\\
                        Per-emb 50  &  $65^{\circ}$ & $25^{\circ}$  & \citet{ValdiviaMena2022}\\
                        IRAS 4A     &  $92^{\circ}$ & $-2^{\circ}$      &  \citet{ValdiviaMena2024}\\
                        G336        &  $75^{\circ}-89^{\circ}$ & $1^{\circ}-15^{\circ}$ & \citet{Olguin2025}\\
                        \bottomrule
                    \end{tabular}
                    }
                \end{threeparttable}
            \label{table:incidentAngles}
        \end{table}
        
    \end{appendix}
    
\end{document}